%% file: banados_RLQ_arxiv1.tex
\documentclass[iop,revtex4]{emulateapj}
\pdfoutput=1

\usepackage{graphicx}
\usepackage{natbib}
\usepackage{amsmath}
\newcommand{\angstrom}{\textup{\AA}}
\newcommand\nar{\ref@jnl{New A Rev.}}
\newcommand{\gps}{\ensuremath{g_{\rm P1}}}
\newcommand{\rps}{\ensuremath{r_{\rm P1}}}
\newcommand{\ips}{\ensuremath{i_{\rm P1}}}
\newcommand{\ipsl}{\ensuremath{i_{\rm P1,lim}}}

\newcommand{\zps}{\ensuremath{z_{\rm P1}}}

\newcommand{\zpsl}{\ensuremath{z_{\rm P1,lim}}}

\newcommand{\yps}{\ensuremath{y_{\rm P1}}}

\newcommand{\PS}{\protect \hbox {Pan-STARRS1}}

\newcommand{\ggrond}{\ensuremath{g_{\rm GROND}}}
\newcommand{\rgrond}{\ensuremath{r_{\rm GROND}}}
\newcommand{\igrond}{\ensuremath{i_{\rm GROND}}}
\newcommand{\zgrond}{\ensuremath{z_{\rm GROND}}}
\newcommand{\Jgrond}{\ensuremath{J_{\rm GROND}}}
\newcommand{\Hgrond}{\ensuremath{H_{\rm GROND}}}
\newcommand{\Kgrond}{\ensuremath{K_{\rm GROND}}}

\newcommand{\intt}{\ensuremath{I_{\rm NTT}}}
\newcommand{\zntt}{\ensuremath{Z_{\rm NTT}}}
\newcommand{\Jntt}{\ensuremath{J_{\rm NTT}}}

\newcommand{\speak}{\ensuremath{S_{\rm 1.4\,GHz,peak}}}

\newcommand{\lya}{Ly\ensuremath{\alpha}}

\newcommand{\Feii}{Fe\ensuremath{\,\textsc{ii}}}

\def\siivpoiv{Si\,{\sc iv}+\,O\,{\sc iv}]}

\def\deg{^\circ}

\def\s{\ifmmode \widetilde \else \~\fi}
\def\={\overline}

\def\spose#1{\hbox to 0pt{#1\hss}}

\def\lta{\mathrel{\spose{\lower 3pt\hbox{$\mathchar"218$}}
     \raise 2.0pt\hbox{$\mathchar"13C$}}}
\def\gta{\mathrel{\spose{\lower 3pt\hbox{$\mathchar"218$}}
     \raise 2.0pt\hbox{$\mathchar"13E$}}}
\def\simlt{\lower.5ex\hbox{\ltsima}}
\def\gtsima{$\; \buildrel > \over \sim \;$}
\def\simgt{\lower.5ex\hbox{\gtsima}}
\shorttitle{The radio-loud fraction of quasars at $z>5.5$}
\shortauthors{Ba\~{n}ados et al.}
\begin{document}

\title{Constraining the radio-loud fraction of quasars at $z>5.5$}

\author{
E.~Ba\~{n}ados\altaffilmark{1},
B.P.~Venemans\altaffilmark{1},
E.~Morganson\altaffilmark{2},
J.~Hodge\altaffilmark{3},
R.~Decarli\altaffilmark{1},
F.~Walter\altaffilmark{1},
D.~Stern\altaffilmark{4},
E.~Schlafly\altaffilmark{1},
E.P.~Farina\altaffilmark{1},
J.~Greiner\altaffilmark{5},
K.C.~Chambers\altaffilmark{6},
X.~Fan\altaffilmark{7},
H-W.~Rix\altaffilmark{1},
W.S. Burgett\altaffilmark{6},
P.W.~Draper\altaffilmark{8},
J.~Flewelling\altaffilmark{6},
N.~Kaiser\altaffilmark{6},
N.~Metcalfe\altaffilmark{8},
J.S.~Morgan\altaffilmark{6},
J.L.~Tonry\altaffilmark{6},
R.J.~Wainscoat\altaffilmark{6}
}

\altaffiltext{1}{Max Planck Institut f\"ur Astronomie, K\"onigstuhl 17, D-69117, Heidelberg, Germany}%
\email{banados@mpia.de}
\altaffiltext{2}{Harvard Smithsonian Center for Astrophysics, 60 Garden St, Cambridge, MA 02138, USA}
\altaffiltext{3}{National Radio Astronomy Observatory, P.O. Box O, Socorro, NM 87801, USA}
\altaffiltext{4}{Jet Propulsion Laboratory, California Institute of Technology, 4800 Oak Grove Drive, Mail Stop 169-221, Pasadena, CA 91109, USA}
\altaffiltext{5}{Max-Planck-Institut f\"ur extraterrestrische Physik, Giessenbachstrasse 1, 85748 Garching, Germany}
\altaffiltext{6}{Institute for Astronomy, University of Hawaii, 2680 Woodlawn Drive, Honolulu, HI 96822, USA}
\altaffiltext{7}{Steward Observatory, The University of Arizona, 933 North Cherry Avenue, Tucson, AZ 85721--0065, USA}
\altaffiltext{8}{Department of Physics, Durham University, South Road, Durham DH1 3LE, UK}

\begin{abstract}
Radio-loud Active Galactic Nuclei at $z\sim2-4$ are typically located in dense environments and
their host galaxies are among the most massive systems at those redshifts, providing key insights
for galaxy evolution. Finding radio-loud quasars at the highest accessible redshifts ($z\sim6$)
is important to study their properties and environments at even earlier cosmic time. They would
also serve as  background sources for radio surveys intended to study the  intergalactic medium
beyond the epoch of reionization in \ion{H}{1} 21 cm absorption. Currently, only five radio-loud
($R=f_{\nu,5{\rm GHz}}/f_{\nu,4400\angstrom}>10$) quasars are known at $z\sim6$. In this paper we
search for $5.5\lesssim z\lesssim7.2$ quasars by cross--matching the optical Pan-STARRS1 and radio
FIRST surveys. The radio information allows identification of quasars missed by typical color-based
selections. While we find no good $6.4\lesssim z\lesssim7.2$ quasar candidates at the sensitivities
of these surveys, we discover two new radio-loud quasars at $z\sim6$. Furthermore, we identify two 
additional $z\sim6$ radio-loud quasars which were not previously known to be radio-loud, nearly 
doubling the current $z\sim6$ sample. We show the importance of having infrared photometry for 
$z>5.5$ quasars to robustly classify them as radio-quiet or radio-loud. Based on this, we reclassify 
the quasar J0203+0012 ($z=5.72$), previously considered radio-loud, to be radio-quiet. Using the 
available data in the literature, we constrain the radio-loud fraction of quasars at $z\sim6$, 
{using the Kaplan--Meier estimator, to be $8.1^{+5.0}_{-3.2}\%$.} This result is consistent with 
there being no evolution of the radio-loud fraction with redshift, in contrast to what has been suggested
by some studies at lower redshifts.
\end{abstract}


\vfil
\eject
\clearpage

\section{INTRODUCTION}
\label{sec:intro}

The study of $z\sim 6$ quasars has shown the presence of almost complete Gunn--Peterson absorption troughs in their spectra, 
corresponding to \lya\ absorption by the neutral Hydrogen in the intergalactic medium (IGM).
This indicates a rapid increase in the neutral fraction of the IGM above $z>6$, providing strong constraints on the end of the epoch of
reionization (EoR) \citep{fan2006c}. The study of the IGM through Gunn--Peterson absorption has its limitations: 
quasar spectra suffer from saturated absorption at $z\gtrsim 6$, and thus it becomes increasingly difficult to study the IGM during 
the EoR \citep[e.g.,][]{furlanetto2006}.

On the other hand, the 21 cm line (unlike the \lya\ line)  does not saturate, allowing the study of the IGM even at large neutral fractions
of Hydrogen \citep[e.g.,][]{carilli2004a}.
Therefore, the identification of radio-loud sources at the highest redshifts will be critical for current and future radio surveys.
These objects will serve as background sources to study the intergalactic medium beyond the EoR through 21 cm absorption
measurements \citep[for example the Low-Frequency Array 
and the Square Kilometer Array; e.g., see ][]{carilli2002}.

Typical high-redshift quasar searches are based on strict optical and near-infrared color criteria chosen to avoid the more numerous cool dwarfs which 
have similar colors to high-redshift quasars \citep[e.g.,][]{fan2001}. An alternative to find elusive quasars with optical colors indistinguishable 
from stars is to require a radio detection. Most of the cool stars that could be confused with high-redshift quasars are not 
radio--bright at mJy sensitivities \citep{kimball2009}; therefore, complementing an optical color-based selection with a bright radio detection reduces the 
contamination significantly \citep[e.g.,][]{mcgreer2009}.

Currently, there are only three $z>5.5$ quasars known with 1.4\,GHz peak flux density $>1$\,mJy (J0836+0054, $z=5.81$, Fan~et~al.~2001; 
J1427+3312, $z=6.12$, McGreer~et~al.~2006, Stern~et~al.~2007; and J1429+5447, $z=6.18$, Willott~et~al.~2010a). 
There are two other $z\sim 6$ quasars in the 
literature classified
as radio-loud, but with fainter radio emission (J2228+0110, $z=5.95$, Zeimann~et~al.~2011; 
and J0203+0012, $z=5.72$, Wang~et~al.~2008).

This small sample of currently identified radio-loud quasars at $z\sim 6$  has already provided important insights into Active Galactic Nuclei (AGN) 
and galaxy evolution, emphasizing the importance of finding more of such objects.
For example, there is evidence that a radio--loud $z\sim 6$ quasar might be located in an overdensity of galaxies \citep{zheng2006}, 
similar to what has been found in radio-loud AGN at lower redshifts \citep[e.g.,][]{wylezalek2013}.
It has also been shown that another $z\sim 6$ radio--loud quasar resides in one of the most powerful known starbursts at $z\sim 6$ \citep{omont2013}. 

At lower redshifts, it is well established that roughly 10\% -- 20\% of all quasars are radio-loud. 
It has been suggested that the radio-loud fraction (RLF) of quasars is a function of both optical luminosity and redshift \citep[e.g.,][]{padovani1993,lafranca1994,hooper1995}.
If a differential evolution between radio-quiet and radio-loud quasars exists, it could indicate changes in properties of black holes such as accretion
modes, black hole masses, or spin \citep{rees1982,wilson1995, laor2000}. This could provide insights into why some quasars have
strong radio emission, while most have only weak radio emission.

Some studies have found evidence of such evolution.
{In particular, \cite{jiang2007} find that the RLF decreases strongly with increasing redshift at a given luminosity. 
For example, they find that the RLF at $M_{2500}= -26$ declines from 24\% to 4\% as redshift increases from 0.5 to 3.
}
\cite{kratzer2014} find a behavior in agreement with these findings
in a similar redshift range ($z \sim 0.5 - 2.5$), but they also point out that the evolution of the RLF closely tracks the apparent
magnitudes, which suggests a possible bias in the results. However, these results are in stark contrast to other studies finding little or no
evidence of such evolution \citep[e.g.,][]{goldschmidt1999,stern2000a,ivezic2002,cirasuolo2003}.

In this paper we take advantage of the large area coverage and photometric information provided by the Faint Images 
of the Radio Sky at Twenty cm survey \citep[FIRST;][]{becker1995} and 
the Panoramic Survey Telescope \& Rapid Response System 1 \citep[\PS, PS1;][]{kaiser2002, kaiser2010} to search for radio-loud quasars 
at $5.5 \lesssim z \lesssim 7.2$. We also revisit the issue of a possible evolution of the RLF of quasars with redshift
by studying the RLF of quasars at the highest accessible redshifts, where an evolution (if existent) 
should be most evident.

The paper is organized as follows. We briefly describe the catalogs used for this work in Section \ref{sec:surveys}. The 
color selection procedures for $5.5 \lesssim z \lesssim 6.4$ and $z\gtrsim 6.4$ quasars with radio counterparts are presented in Section \ref{sec:selection}.
In Section~4, we present our follow-up campaign and the discovery of two new radio-loud
 $z\gtrsim 5.5$ quasars. 
  The radio-loud definition used in this paper and details on how it is calculated are introduced in Section~\ref{sec:RLD}.
 In Section~\ref{sec:results}, we investigate the radio-loud fraction of $z > 5.5$
quasars by compiling radio information on all such quasars currently
in the literature.  This work identifies two additional high-redshift,
radio-loud quasars that had not previously been noted to be radio-loud.

We summarize our results in Section \ref{sec:summary}. Magnitudes throughout the paper are given in the AB system.
We employ a cosmology with $H_0 = 69.3 \,\mbox{km s}^{-1}$ Mpc$^{-1}$, $\Omega_M = 0.29$, and $\Omega_\Lambda = 0.71$ \citep[][]{wmap9}.

\section{Survey Data}
\label{sec:surveys}

\subsection{FIRST}
The FIRST survey was designed to observe the sky at 20 cm (1.4 GHz) matching a region
of the sky mapped by the Sloan Digital Sky Survey \citep[SDSS;][]{york2000}, covering a total of about $10,600$ square degrees. The survey 
contains more than $900,000$ unique sources, with positional accuracy to $\lesssim 1\arcsec$. The catalog has a 5$\sigma$ detection threshold which
typically corresponds to 1\,mJy although
there is a deeper equatorial region where the detection threshold is about 0.75\,mJy.

\subsection{Pan-STARRS1}

The PS1 $3\pi$ survey
has mapped all the sky above declination $-30\deg$ over a period of $\sim 3$ years in five optical filters \gps, 
\rps, \ips, \zps, and \yps\ \citep{stubbs2010, tonry2012}. The PS1 catalog used in this work comes from the
 first internal release of the $3\pi$ stacked catalog (PV1), which is based on the co-added 
PS1 exposures \citep[see][]{metcalfe2013}. This catalog includes data obtained primarily during the 
period 2010 May--2013 March and the stacked images consist on average of the co-addition of $\sim 8$ single
images per filter. The $5\sigma$ median limiting magnitudes of this catalog are
$\gps=22.9$, $\rps=22.8$, $\ips=22.6$, $\zps=21.9$, and $\yps=20.9$.
PS1 goes significantly deeper than SDSS in the $i$ and $z$ bands which together with 
the inclusion of a near-infrared $y$ band 
allow it to identify new
high-redshift quasars even in areas already covered by SDSS.

\section{Candidate Selection}

\label{sec:selection}

\subsection{The FIRST/Pan-STARRS1 catalog}

In this paper we use the multiwavelength information and large area coverage of the FIRST and Pan-STARRS1 surveys to 
find radio-loud high-redshift quasars. 
We cross match the FIRST catalog (13Jun05 version) and the PV1 PS1 stack catalog using a 2\arcsec\ matching radius. 
This yields a catalog containing $334,290$ objects.
{Given the similar astrometric accuracy of Pan-STARRS1 and SDSS, we use the same matching radius 
utilized by the SDSS spectroscopic target quasar selection \citep{richards2002b}. The peak of the SDSS-FIRST positional offsets
occurs at $\sim 0.2\arcsec$ and the fraction of false matches within 2\arcsec\ is about 0.1\% \citep[][see their Figure~6]{schneider2007,schneider2010}.
 Although this matching radius introduces a bias against quasars with extended radio morphologies 
(e.g., double-lobe quasars without radio cores or lobe-dominated quasars), \cite{ivezic2002} show
that less than 10\% of SDSS-FIRST quasars have complex radio morphologies. 
}

As redshift increases, the amount of  neutral hydrogen in the universe also increases. At $z \gtrsim 6$ the optically thick Ly$\alpha$ forest absorbs most of the 
light coming from wavelengths $\lambda_{{\rm rest}} < 1216$ \AA. This implies that objects at  $z\sim 7$ ($z\sim 6$) are undetected 
or very faint in the $z$-band ($i$-band), showing a `drop' in their spectra. They are thus called $z$-dropouts ($i$-dropouts). This `drop' can be measured
by their red  $z-y$ and $i-z$ colors for $z$-dropouts and $i$-dropouts, respectively.

Given that the radio detection requirement significantly decreases the amount of contaminants (especially cool dwarfs), 
we perform a much broader selection criteria in terms of colors and signal-to-noise ratio (S/N) in comparison with 
our $z$- and $i$-dropout criteria presented in \cite{venemans2015} and \cite{banados2014}, respectively.
{
In particular for the $z$-dropouts in this paper, we allow for objects undetected in the $\zps$ band to be 0.1 mag bluer than
in \cite{venemans2015} and relax the S/N criteria in their \gps\ and \rps\ bands to be $\mathrm{S/N} < 5$ instead of $\mathrm{S/N} < 3$ 
(see Section \ref{sec:zdrop}).
For the $i$-dropouts in this paper, we relax the selection limits in comparison with \cite{banados2014}. 
We allow the candidates to be 0.7 mag and 0.4 mag bluer in the $\ips -\zps$ and $\rps -\zps$ colors, respectively.
For this selection we do not put any constrain in $\zps - \yps$ color (see Section \ref{sec:idrop}).
}
In this way we can detect $z>5.5$ quasars which have similar colors to cool stars that are missed by typical color-based criteria.

\subsection{$z$-dropout  catalog search ($z\gtrsim 6.4$)}
\label{sec:zdrop}

We require that more than 85\% of the expected point-spread function (PSF)-weighted
flux in the \zps\ and the \yps\ bands is located in valid pixels 
(i.e., that the PS1 catalog entry has
{\tt PSF\_QF} $> 0.85$). We require a S/N $>7$ in the \yps\ band and exclude 
those measurements in the \yps\ band flagged as suspicious by the 
Image Processing Pipeline \citep[IPP;][]{magnier2006,magnier2007} (see Table~6 in Ba\~nados et al. 2014).
The catalog selection can be summarized as:
\begin{subequations}
\begin{eqnarray}
\mbox{S/N}(\yps) > 7 & \label{eq:ysn} \\
((\mbox{S/N}(\zps) \geq 3 )\; \mbox{AND} \, (\zps - \yps > 1.4)) \; \mbox{OR} & \nonumber \\
((\mbox{S/N}(\zps) < 3 ) \;  \mbox{AND}\, (\zpsl - \yps > 1.3)) & \label{eq:zycolor} \\
(\mbox{S/N}(\ips) < 5) \; \mbox{OR} \; (\ips - \yps > 2.0) & \label{eq:iycolor}\\
(\mbox{S/N}(\rps) < 5)  &\label{eq:rcolor} \\
(\mbox{S/N}(\gps) < 5) &\label{eq:gcolor} 
\end{eqnarray}
\end{subequations}
where \zpsl\ is the 3$\sigma$ \zps\ limiting magnitude. This selection yields 66 candidates.

\subsection{$i$-dropout catalog search  ($5.5 \lesssim z \lesssim 6.4$)}
\label{sec:idrop}

Similar to the $z$-dropout catalog search, we require that more than 85\% of 
the expected PSF-weighted flux in the
\ips\ and the \zps\ bands is located in valid pixels. We require
a S/N $>10$ in the \zps\ band and exclude those measurements flagged as suspicious
by the IPP in the \zps\ band.

We do not put any constraint on the \yps\ band. This allows us to identify 
quasar candidates across a broad redshift range ($5.5 \lesssim z \lesssim 6.4$) and
make better use of the \zps\ band depth (which is deeper than the \yps\ band).
The \yps\ information is used later on for the follow-up campaign.

We can summarize the catalog selection criteria as follows:
\begin{subequations}
\begin{eqnarray}
\mbox{S/N}(\zps) >  10& \label{eq:zsn} \\
((\mbox{S/N}(\ips) \geq 3 )\; \mbox{AND} \, (\ips - \zps > 1.5)) \; \mbox{OR} & \nonumber \\
((\mbox{S/N}(\ips) < 3 ) \;  \mbox{AND}\, (\ipsl - \zps > 1.0)) & \label{eq:izcolor} \\
(\mbox{S/N}(\rps) < 3) \; \mbox{OR} \; (\rps - \zps > 1.8) & \label{eq:rzcolor} \\
(\mbox{S/N}(\gps) < 3) \; \mbox{OR} \;(\gps - \zps > 1.8) & \label{eq:gcolor1}  
\end{eqnarray}
\end{subequations}
where \ipsl\ is the 3$\sigma$ \ips\ limiting magnitude. 
This query yields 71 candidates.

\subsection{Visual Inspection}
\label{ref:visual}

The number of candidates obtained from Sections \ref{sec:zdrop} and  \ref{sec:idrop} are 
small enough to visually inspect all of them. 
We use the latest PS1 images available for the visual inspection, which
are usually deeper than the images used to generate the PS1 PV1 catalog.
We also perform forced photometry on them to corroborate the catalog colors
(as described in Ba\~nados et al. 2014), especially when PV1 only reports limiting 
magnitudes.
Thus, we visually inspect all the PS1 stacked, FIRST, 
and \zps\ and \yps\ PS1 single epoch images (where the S/N is expected to be the
highest) for every candidate.
{The most common cases eliminated by visually inspection are PS1 artifacts due to some bad single epoch images, objects
that lacked information in the PV1 catalog, and objects with evident extended morphology in the optical images.}
Based on the visual inspection,  we assign priorities for the follow-up. 
Low-priority candidates are the ones whose 
PS1 detections look questionable, in the limit of our S/N cut,{ and/or objects with extended radio morphology which produces
slightly positional offsets ($\gtrsim 1\arcsec$) between the optical and radio sources.
}

\subsubsection{$z$-dropouts}

Almost all the candidates can be ruled out by their PS1 stack images and/or single epoch images.
There are only two objects we cannot completely rule out
{although there are some lines of evidence pointing us to believe they are unlikely to be quasars.
For both PSO~J141.7159+59.5142 and PSO~J172.3556+18.7734, the \yps\ detection looks questionable, in the limit of our S/N cut: 7.5 and 7.0, 
respectively. Their \yps\ catalog aperture and PSF magnitudes differ by 0.3 and 0.28 mag  which could indicate they are 
extended sources but it is hard to tell at this low S/N. The optical and radio positional offsets are somewhat larger than for most
of the candidates: $0.6\arcsec$ and $1.5\arcsec$. All of this combined makes them low priority candidates.
}
The PS1 and FIRST information for these sources is listed in Table~\ref{tab:candidates}.

\subsubsection{$i$-dropouts}
One of the candidates we selected is the known radio-loud quasar J0836+0054 at $z=5.82$ \citep{fan2001}. Its images look good and we would have followed it up.
There are two low-redshift quasars that could have been selected for follow-up, J0927+0203 a quasar with a bright H$\alpha$ line at
$\sim 9200$\AA\  \citep[$z=0.39$;][]{schneider2010} and J0943+5417, an \Feii\ low-ionization broad absorption line quasar  \citep[$z=2.22$;][]{urrutia2009}.
After the visual inspection and literature search,
there are 10 remaining candidates,{ out of which 9 are high priority candidates}. Their PS1 and FIRST photometry are listed in Table~\ref{tab:candidates}.

\section{Follow-up}

\subsection{Imaging}
We use a variety of telescopes and instruments to confirm the optical colors and to obtain near-infrared photometry of our candidates, 
thereby allowing efficient removal of interlopers.

GROND \citep{greiner2008} at the 2.2m telescope in La Silla was used to take simultaneous images in the filters $grizJHK$ during 2014 January 24 -- February 5.
Typical on-source exposure times were 1440 s in the near-infrared and 1380 s in the optical.
The ESO Faint Object Spectrograph and Camera 2  \citep[EFOSC2;][]{buzzoni1984} and the infrared spectrograph and imaging camera
Son of ISAAC \citep[SofI;][]{moorwood1998} at the ESO New Technology Telescope (NTT) were used to perform 
imaging in the \intt\ ($i\# 705$), \zntt\ ($z\#623$), and \Jntt\ bands during 2014 March 2--6 with on-source exposure times of 600 s in the \intt\ and \zntt\ bands and 300 s in the 
\Jntt\ band. 
The data reduction consisted of bias subtraction, flat fielding, sky subtraction, image alignment, and stacking. The photometric zeropoints were determined
as in \cite{banados2014}\footnote{Color conversions missing in \cite{banados2014}:\\ $\ggrond =  \gps + 0.332 \times (\gps - \rps) + 0.055$; \\
$\rgrond =  \rps + 0.044 \times (\rps - \ips) - 0.001$; and \Jntt\ and \Kgrond\ are calibrated against 2MASS.} and their errors are included in the magnitudes reported in this work. All of our high-priority candidates were photometrically 
followed up except for one which we directly observed spectroscopically (see next Section). Two low-priority $z$-dropouts and one low-priority $i$-dropout 
are still awaiting follow-up.
Table~\ref{tab:followup} shows the follow-up photometry of our candidates.

\subsection{Spectroscopy}
\label{sec:spectroscopy}
We have taken spectra of four high-priority objects that were not rejected by the follow-up photometry. We processed the data using standard techniques, including bias subtraction, flat fielding, 
sky subtraction,  combination of individual frames, wavelength calibration, and spectrum extraction. The spectra were flux calibrated using 
standard stars from \cite{massey1990} and \cite{hamuy1992, hamuy1994}.

The candidate PSO J114.6345+25.6724 was observed with the FOcal Reducer/low dispersion Spectrograph 2 \citep[FORS2;][]{appenzeller1992} at
the Very Large Telescope (VLT) on 2014 April
26 with an exposure time of 1497~s. The spectrum shows no clear break in the continuum and we classify this object as a radio galaxy
at $z=1.17$ by the identification of the narrow [O$\,${\sc ii}] $\lambda 3728$ emission line.

We obtained an optical spectrum of the candidate PSO~J354.6110+04.9453
using the Double Spectrograph on the 5~m Hale telescope at Palomar Observatory (DBSP) on 2014 July 21 for a total integration time of 1800~s. 
The object has a red spectrum lacking the clear (\lya) break which is a typical signature of high-redshift quasars. The spectrum does not 
clearly identify lines to determine 
a redshift. We believe this object is most likely a radio galaxy, but it is definitely not a $z>5.5$ quasar. 
The other two spectroscopically followed up objects -- PSO~J055.4244--00.8035 and PSO~J135.3860+16.2518 --
were confirmed to be high-redshift quasars.
{These two newly discovered quasars would not have been selected as candidates by the optical selection
criteria presented in \cite{banados2014} (although PSO~J055.4244--00.8035 has only a lower limit of $\ips - \zps > 1.3$ and it might have been selected
if deeper $\ips$ data was available). The observations for these quasars are outlined in more detail below.
}

\subsubsection{PSO J055.4244--00.8035 ($z=5.68 \pm 0.05$)} 
\label{sec:P055-00}
The discovery spectrum was taken on 2014 February 22 using the DBSP spectrograph with a total exposure time of 2400~s.
These observations were carried out in $\sim 1\arcsec$ seeing using the 1\farcs5 wide longslit.
This spectrum shows a sharp \lya\ break indicating that the object is unambiguously a quasar at $z > 5.5$ but the S/N does not allow us to determine
an accurate redshift. We took a second spectrum with FORS2 at the VLT on 2014 August 4; the seeing was 1\farcs1 and it was observed for 1467~s.
This spectrum is shown in Figure \ref{fig:spectra} and there are no obvious lines to fit and use to derive a redshift.
{There is, however, a tentative \siivpoiv\ line which falls in a region with considerable sky emission and
telluric absorption making it not reliable for redshift estimation.}
We estimate the redshift instead by comparing the observed quasar spectrum with the composite SDSS $z\sim 6$ quasar spectrum
from \cite{fan2006a}.  We assume the redshift that minimizes the $\chi^2$ between the observed spectrum and the template
(the wavelength range where the minimization is performed is $\lambda_{\rm rest}$ = 1240 -- 1450 \AA).
The estimated redshift is $z=5.68$. However, because of the lack 
of strong features in the spectrum, the $\chi^2$ distribution is relatively flat around the minimum 
and thus a range of redshifts is acceptable.
We follow \cite{banados2014} and assume a redshift uncertainty of 0.05.

\subsubsection{PSO J135.3860+16.2518 ($z=5.63 \pm 0.05$)}
\begin{figure}[ht]
\begin{center}
\includegraphics[scale=0.55]{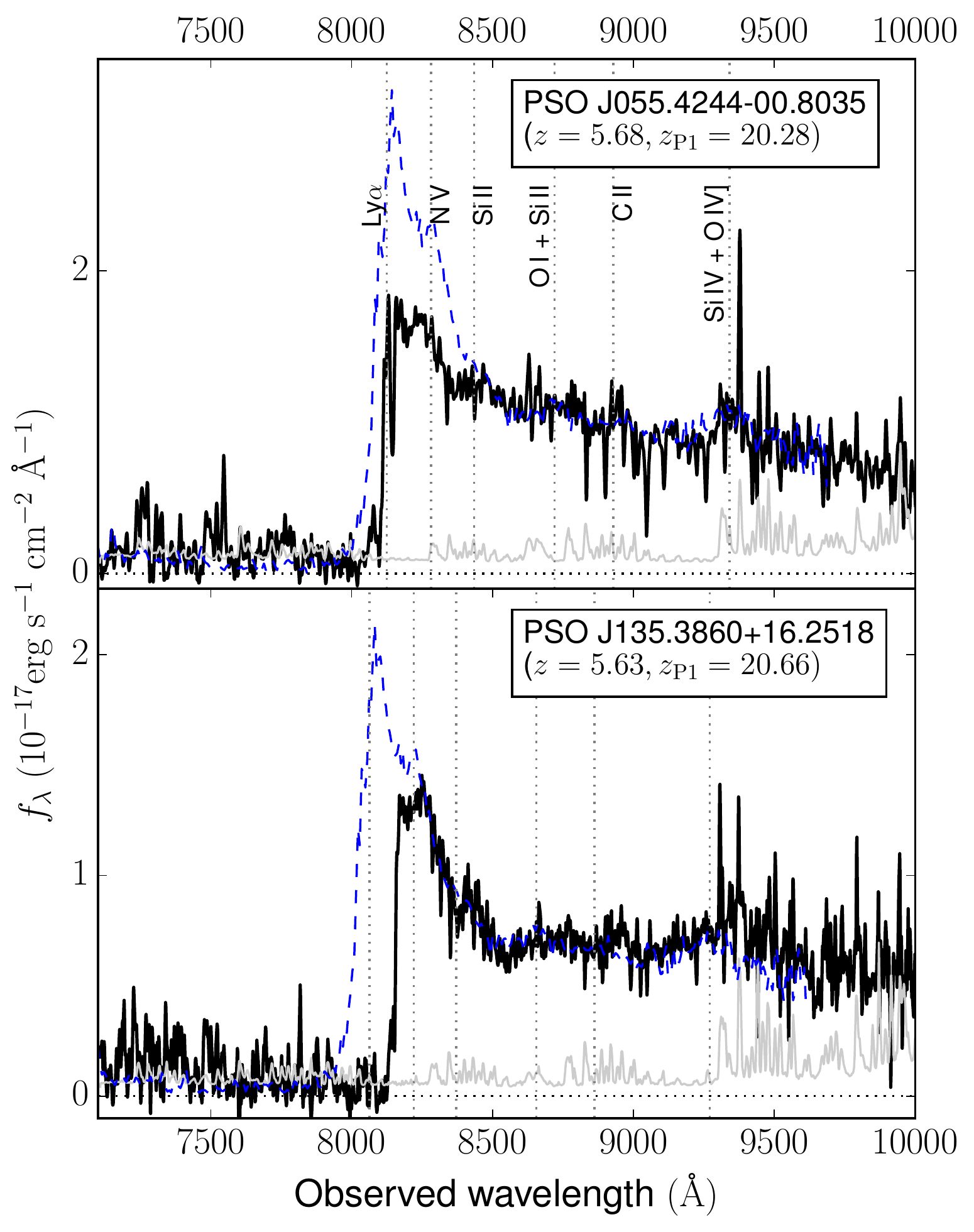}
 \caption{Spectra of the two radio-loud quasars discovered in this paper. The gray solid lines show the 1$\sigma$ error in the spectra.
 The blue dashed line is the SDSS $z\sim6$ composite quasar spectrum from \cite{fan2006a} at the redshift of the quasars for comparison.
 Vertical dotted lines indicate the observed wavelengths of key spectral lines, as given in the top panel. 
 \label{fig:spectra}}
 \end{center}
 \end{figure}

The discovery spectrum was taken with  EFOSC2 at the NTT on 2014 March 3. The observations were carried out with the Gr\#16 grism, 1\farcs5 slit width, 1\farcs3 seeing, 
and a total exposure time of 3600~s.  
The spectrum is very noisy but it resembles the shape of a high-redshift quasar with a tentative 
\lya\ line at $\sim 8100\, $\AA. In order to increase the S/N and confirm the quasar redshift, we took two additional spectra and combined them.
One spectrum was taken on 2014 April 5 with the Multi-Object Double Spectrograph \citep[MODS;][]{pogge2010} at the
Large Binocular Telescope (LBT). The observations were carried out under suboptimal weather conditions with the 1\farcs2 wide longslit
for a total exposure time of 2400~s. The second spectrum was taken with FORS2 at the VLT on 2014 April 26 with a total exposure time of 1467~s.
The observing conditions
were excellent with 0\farcs55 seeing and we used the 1\farcs3 width longslit. 
The combined MODS-FORS2 spectrum is shown in Figure \ref{fig:spectra}. The estimated redshift by matching to
the $z\sim6$ quasar composite spectrum from \cite{fan2006a} 
is $z=5.63 \pm 0.05$. 
{ 
The redshift estimate for PSO~J135.3860+16.2518 is quite uncertain as represented by its error bar. There might be 
 a tentative \siivpoiv\ line which would place this quasar at a slightly higher redshift. However, this line falls in the same 
 region as the tentative line in the quasar PSO~J055.4244--00.8035, which is not a reliable region for redshift determination.
 A higher S/N optical spectrum and/or a near-infrared spectrum would be beneficial to obtain a more accurate redshift.
}
 
\section{Radio-Loudness}

\label{sec:RLD}
A clear consensus on a boundary between radio-loud and radio-quiet quasars has been difficult to achieve and there are several radio-loudness 
criteria in the literature
\citep[for a comparison of the different criteria see][]{hao2014b}. We adopt the most widespread definition 
in the literature. This is the radio/optical flux density ratio,
$R=f_{\nu,5 {\rm GHz}} / f_{\nu,4400 \angstrom} $  \citep{kellermann1989}, where $f_{\nu,5 {\rm GHz}}$ is the 5\,GHz radio rest-frame flux density,
$f_{\nu,4400 \angstrom}$ 
is the 4400 \AA\ optical rest-frame flux density, and a quasar is considered radio loud if $R>10$.

\subsection{The Radio Emission}
The rest-frame $5$\,GHz radio flux density is obtained from the observed peak flux density at $1.4$\,GHz.
We use the peak flux density since most of the $z\sim 6$ quasars appear to be unresolved on the radio maps \citep[e.g., ][but see also \citealp{cao2014}, where 
they claim that there
may be extended structures around the radio-loud quasar J2228+0110 on arcsecond scales]{wang2007, wang2011}.

We assume a power-law ($f_\nu \sim \nu^\alpha$)
radio spectral energy distribution, adopting a typical radio spectral index $\alpha_R = -0.75$ as used in other high-redshift studies \citep[e.g.,][]{wang2007, momjian2014}. This 
index appears to be appropriate as \cite{frey2005, frey2008, frey2011} find that at least three $z\sim 6$ radio-loud quasars 
show a steep radio spectrum. 

\subsection{The Optical Emission}
\label{sec:optical}
The optical spectral indices of quasars span a fairly large range (at least $ -1 < \alpha_\nu < 0$).
When a direct measurement of the optical rest-frame flux density at 4400\,\AA\ is not possible, this is typically 
extrapolated from the AB magnitude at rest frame 1450\,\AA\ ($m_{1450}$) assuming an average optical spectral index of $\alpha_\nu = -0.5$ 
\citep[e.g.,][]{wang2007}. These fairly large extrapolations could lead to dramatic errors if the studied quasars are not average quasars.
At $z \sim 6 - 7$ we can take advantage of infrared space missions such as \textit{Spitzer} and the \textit{Wide-Field Infrared Survey}
\cite[\textit{WISE};][]{wright2010}
to reduce the extrapolation error by using $\sim 3\,\mu$m ($\lambda_{\rm rest} = 4286\angstrom$ at $z=6$) 
 photometry.

 Following previous works, we also assume an optical spectral index of $\alpha_\nu = -0.5$ but we estimate $f_{\nu,4400 \angstrom}$ from  IRAC 
 \citep{fazio2004} $3.6 \mu$m
  ($S_{3.6\,\mu m}$) observations when available ($\lambda_{\rm rest, eff}=5071$\AA\ at $z=6$). Otherwise, $f_{\nu,4400 \angstrom}$ is calculated from the \textit{WISE} $W_1$ magnitude
($\lambda_{\rm rest, eff}=4811$\AA\ at $z=6$) reported in the ALLWISE Source catalog or reject table (with $S/N > 2.5$). 
 Table~\ref{tab:literature} lists $S_{3.6\,\mu\rm{m}}$, $W_1$, and $m_{1450}$ for all $z>5.5$ quasars with published measurements at
$1.4$\,GHz in the literature.
There are five quasars without IRAC or \textit{WISE} measurements. 
For these objects, $f_{\nu,4400 \angstrom}$ is estimated from $m_{1450}$. To estimate the error we determine $f_{\nu,4400 \angstrom}$ from $m_{1450}$
for the quasars having IRAC or \textit{WISE} ($\sim 3\,\mu$m) measurements. 
{ Then, for each object we compute the ratio
$(f_{\nu,4400 \angstrom}(3\,\mu \rm{m}) - f_{\nu,4400 \angstrom}(m_{1450})) / f_{\nu,4400 \angstrom}(3 \,\mu \rm{m})$. This results in a
symmetric distribution, centered on zero, and with a standard deviation of 0.4. Finally, 
we take the absolute value of the previous distribution and assume its median as a representative error
for $f_{\nu,4400 \angstrom}$ derived from $m_{1450}$. This corresponds to a relative error of 0.30
(i.e., analogous to assuming a measurement of $f_{\nu,4400 \angstrom}$ with $S/N = 3.3$).
}
These errors must be taken with caution since they are just representative uncertainties and there could be objects with considerably
larger errors, as exemplified in Section \ref{sec:j0203}.

\section{Results}
\label{sec:results}
We determine the radio-loudness parameter $R$ and the optical and radio luminosities $L_{4400\angstrom}$ and $L_{\rm 5GHz}$ ($L = \nu L_\nu$)
for all the $z>5.5$ quasars having 1.4\,GHz data published in the literature. These parameters are included in Table~\ref{tab:literature}.
Figure~\ref{fig:radio} shows the rest-frame 5\,GHz radio luminosity versus rest-frame 4400\,\AA\ optical
luminosity. Eight quasars are classified as radio-loud ($R>10$),
including the two quasars discovered {in Section \ref{sec:spectroscopy} and two additional possible radio-loud 
quasars which will be introduced in Section \ref{sec:newfirst}.
There are 33 objects robustly classified as radio-quiet quasars ($R<10$). Two quasars 
need deeper radio data to classify them unambiguously (with radio loudness upper limit $>10$)}.  In this paper we do not find any radio-loud quasar at $z \sim 7$ in an area of about $10,600$ square
degrees of sky to the sensitivities of the FIRST and \PS\ surveys ($\sim 1$\,mJy and $7\sigma$-limiting magnitude $\yps\sim 20.5$,
respectively). Conclusions on the RLF at $z>6.5$ are not possible at this time, since 
there are currently only {seven quasars known at $z>6.5$  \citep{mortlock2011, venemans2013, venemans2015}}. 
{From these seven quasars, only one has dedicated radio follow-up \citep{momjian2014} while a second one is 
in the FIRST footprint but it is a non-detection \citep[see Section \ref{sec:rlfz6};][]{venemans2015}}.  
The RLF at $z\sim 6$ is discussed in Section~\ref{sec:rlfz6}.

\subsection{J0203+0012: A Radio-Loud Quasar?}
\label{sec:j0203}
The quasar J0203+0012 at $z=5.72$ was classified as radio-loud quasar by \cite{wang2008b}. They derived the rest-frame 4400 \AA\ flux density from 
the 1450\,\AA\  magnitude. As discussed previously, these large extrapolations can carry large uncertainties and can be critical for the classification
and derived parameters for specific objects. This is the case for J0203+0012, a broad-absorption line quasar \citep{mortlock2009} whose 
$m_{1450}$ could have been underestimated, resulting in a low optical luminosity (see also its spectral energy distribution 
in Figure~14 of \citealp{leipski2014}).
This quasar has a radio-loudness parameter $R=4.3\pm 0.5$ or $R=12.1\pm 3.2$ depending if the rest-frame 4400 \AA\ flux density
is extrapolated from the IRAC 3.6\,$\mu$m photometry or from $m_{1450}$, respectively (see Figure~\ref{fig:radio}).
It is clear that the difference is dramatic and that
by using the $m_{1450}$ proxy this quasar would be classified (barely) as radio-loud. We argue that the value of $L_{4400\angstrom}$  obtained from the 
observed $3.6 \,\mu$m photometry
is more reliable for $z>5.5$ quasars since it relies less on
extrapolation (for this particular case, the extrapolation is less by approximately a factor of  three).
Also, while radio-loud AGN are typically located
in dense environments \citep[e.g.,][]{venemans2007b,hatch2014}, we found that J0203+0012 does not live in a particularly dense
region but rather comparable with what is expected in blank fields \citep{banados2013}.

\subsection{Pushing the FIRST Detection Threshold}
\label{sec:newfirst}

The FIRST survey has a typical source detection threshold of 1 mJy\,beam$^{-1}$ which assures that the catalog has reliable entries with typical S/N
greater than five.
There are 30 known $z>5.5$ quasars that are not in the FIRST catalog or in the Stripe 82 VLA Survey catalog \citep{hodge2011} but that are located in the FIRST footprint.
The discovery papers of these quasars are
\cite{mahabal2005, cool2006, jiang2009, willott2009, willott2010b, willott2010a, mcgreer2013, banados2014, venemans2015}; 
Ba\~nados et al. (in prep.); Venemans et al. (in prep.); and Warren et al.~(in prep.).

We checked for a radio detection beyond the FIRST catalog threshold as follows: 
we obtained the 1.4\,GHz FIRST images for all 30 quasars and checked for radio emission within 3\arcsec\ of the optical quasar position
with a S/N$\geq 3$. We find that the quasars J1609+3041 at $z=6.14$ (Warren et al. in prep) and J2053+0047 at $z=5.92$ \citep{jiang2009}
 have tentative 1.4\,GHz detections at S/N of 3.5 and 3, respectively. Their radio postage stamps are shown in Figure~\ref{fig:firstqsos}.
 The two quasars have optical-to-radio positional differences less than 1\farcs8 (1 pixel).
 {In order to quantify the probability of finding a spurious association with a $3\sigma$ fluctuation given our sample of 30 quasars
 we performed the  following steps. We placed 100 random positions in each 1 arcmin$^2$ FIRST image centered on a quasar and measured
 the maximum peak flux within 1\farcs8. We removed points falling within 1\farcs8 from an optical source in the PS1 catalog. 
 There are no radio sources in the FIRST catalog for any of these 1 arcmin$^2$ fields. We computed the fraction of measurements
 with S/N$\,\geq 3$. We repeated this procedure 100 times and the fraction of measurements with S/N$\,\geq 3$ was always $< 1 \%$.
 The full distribution is centered on $0.5\%$ with a standard deviation of $0.1\%$. Therefore, in our sample of 30 quasars, the expected
 number of spurious $\sim 3\sigma$ associations within 1\farcs8 is $0.15$ and being conservative less than $0.3$. This analysis
suggests that these identifications are unlikely to be spurious.
 }
 If these detections are real, these quasars are classified as radio-loud with $R=28.3\pm 8.6$ and $R=44.1\pm18.7$ (see Table~\ref{tab:literature}). 
In Figure~\ref{fig:radio}, these new tentative radio detections are marked as red downward triangles.

We make a mean stack of the 28 remaining quasars in the FIRST footprint that have $\mathrm{S/N}<3$ in their individual images.
We find no detection in the stacked image with an upper limit of $f_\nu=84\, \mu$Jy (see Figure~\ref{fig:radio}).

\subsection{Constraining the Radio-Loud Fraction of Quasars at $z\sim 6$}
\label{sec:rlfz6}
Considering all the quasars in Table~\ref{tab:literature}, there
are eight known radio-loud quasars at $z\sim 6$, 32 radio-quiet (excluding J1120+0641 at $z=7.08$), and two ambiguous.
{There is one additional quasar that is robustly classified by a non-detection in FIRST as radio--quiet: 
J0148+0600 at $z=5.98$ 
\citep{banados2014, beckerG2015}. This radio-quiet quasar has $\log L_{4400\angstrom}\, (L_\sun) = 13.04 \pm 0.2 $,  
$\log L_{\rm 5\,Ghz}(L_\sun) < 8.7 $, and $R<5.6$  (see Figure~\ref{fig:radio}).}
We can provide a rough estimation of the radio-loud fraction of
quasars at $z \sim 6$ of RLF\,$= 8 / (8 + (34 + 1)) \sim 19\%$. In this statistics we considered the two ambiguous
cases as radio-quiet. 
This is a relatively large fraction; however, these quasars were selected by several methods
which could potentially bias the results. As we have included radio-loud
quasars that could not have been discovered based on their optical/near-infrared properties alone, the actual fraction of radio-loud quasars is overestimated.
Therefore, this value has to be taken only as an upper limit.

In order to set a lower limit in the RLF at $z\sim 6$, we consider quasars that were selected based on their optical properties 
(i.e., we exclude the two quasars discovered in this paper, J2228+0110 which was discovered by its radio emission by Zeimann et al. 2011, 
and J1427+3312 which was discovered by its radio emission by McGreer et al. 2006\footnote{We note however, that J1427+3312 was independently discovered
 by \cite{stern2007} without using the radio information, but  using a mid-infrared selection instead.}).
 {We also exclude quasars at $z>6.5$, i.e., J1120+0641 at $z=7.08$ and PSO~J036.5078+03.0498 at $z=6.527$. 
 The latter quasar was discovered in \cite{venemans2015}. It is not detected in FIRST and  
 has  $\log L_{4400\angstrom}\, (L_\sun) = 12.87 \pm 0.03 $,
$\log L_{\rm 5\,Ghz}(L_\sun) < 8.8 $, and $R<10.7$  (see Figure~\ref{fig:radio}).
}
Considering all the FIRST non-detections from the previous section as radio-quiet,
we find a lower limit of RLF $= 4 / (4 + (34 + 27)) \sim  6\%$. This is a lower limit because there is still the possibility that a fraction of
the FIRST non-detections are radio-loud (see Figure~\ref{fig:radio}). Therefore, in this case we are potentially 
 underestimating the number of radio-loud quasars.
 
{
 Additionally, in order to fully use the information provided by both the radio detections and upper limits, 
 we estimate the radio-loud fraction using the Kaplan--Meier estimator \citep{kaplan1958}.
The RLF estimated with this method, after excluding quasars at $z>6.5$ and quasars selected by their radio emission, is $8.1^{+5.0}_{-3.2}\%$.
 }

\subsection{What Changes with an Alternative Radio-Loudness Definition?}
\label{sec:alternative}
\begin{figure*}[ht]
\begin{center}
\includegraphics[scale=0.9]{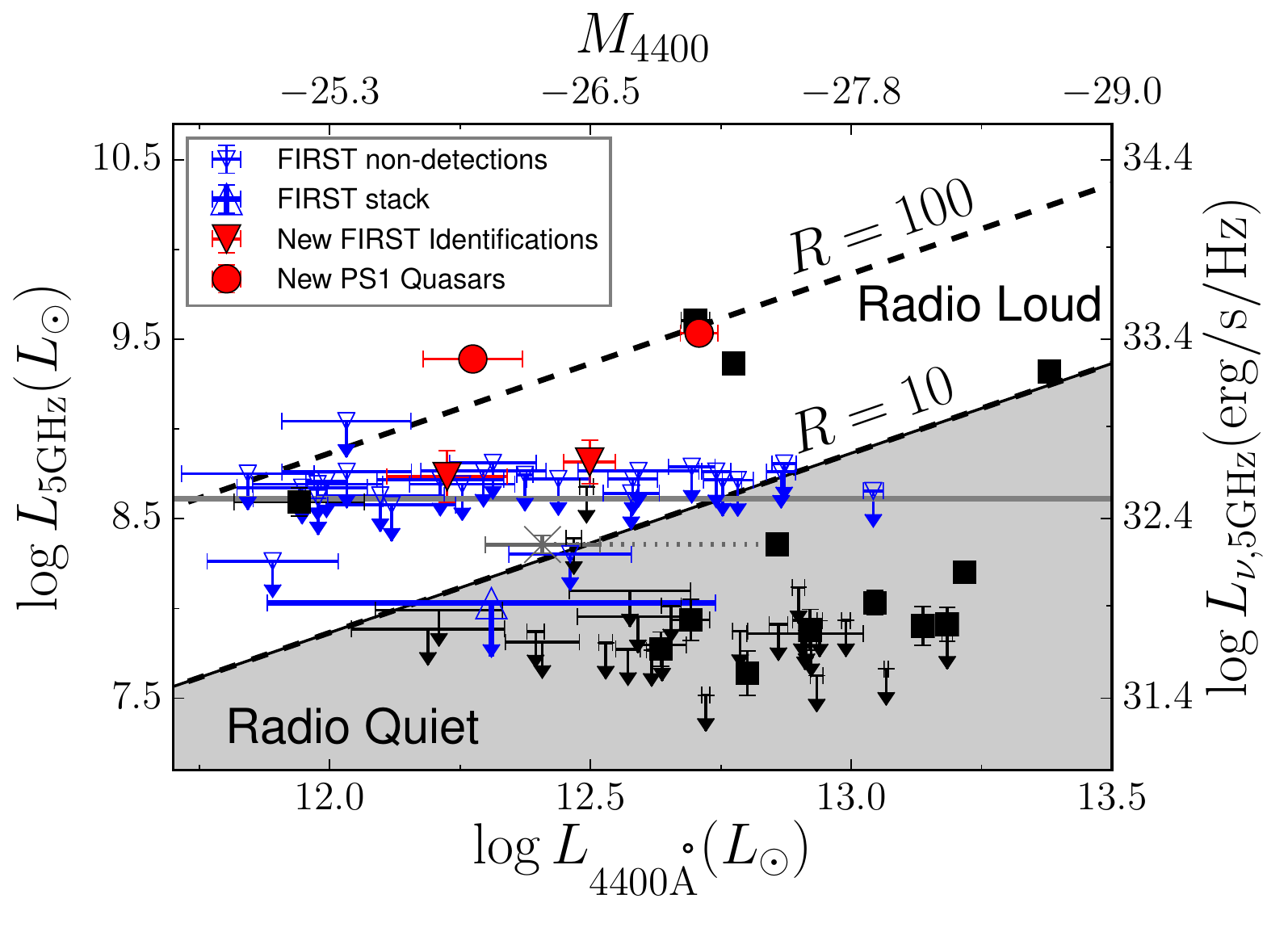}
 \caption{Rest-frame 5\,GHz radio luminosity versus rest-frame $4400$\AA\ optical luminosity for all $z>5.5$ quasars
 in the literature observed at 1.4\,GHz (see Table~\ref{tab:literature}).
 Filled symbols are 1.4\,GHz detections and arrows represent $3\sigma$ upper limits in the rest-frame 5GHz luminosity.
 The dashed lines represent the radio-to-optical ratios ($R$) of 100 and 10 (separation of radio-loud and radio-quiet). The red circles
 are the two new \PS\ radio-loud quasars presented here. These quasars are among the most radio loud quasars at these redshifts.
 The red downward triangles are two quasars that we identify as radio-loud by inspecting their FIRST images (see Section~\ref{sec:newfirst}).
 The blue downward triangles are upper limits for 28 quasars that are located in the FIRST survey footprint. The blue upward triangle represents the location
 of the stack of the 28 undetected quasars in the FIRST survey assuming their average redshift of 6.
 The gray cross represents the position of J0203+0012 if its $L_{4400\angstrom}$ is calculated from extrapolating $m_{1450}$ and it is 
 connected by a  dotted line with the position determined by estimating $L_{4400\angstrom}$ from the IRAC $3.6\,\mu$m photometry (see Section \ref{sec:j0203}).
 The horizontal gray line denotes the division of radio-loud and radio-quiet quasars employed by \cite{jiang2007} (see Section~\ref{sec:alternative}).}
 \label{fig:radio}
 \end{center}
 \end{figure*}

\begin{figure*}[ht]
\begin{center}
\includegraphics[scale=0.9]{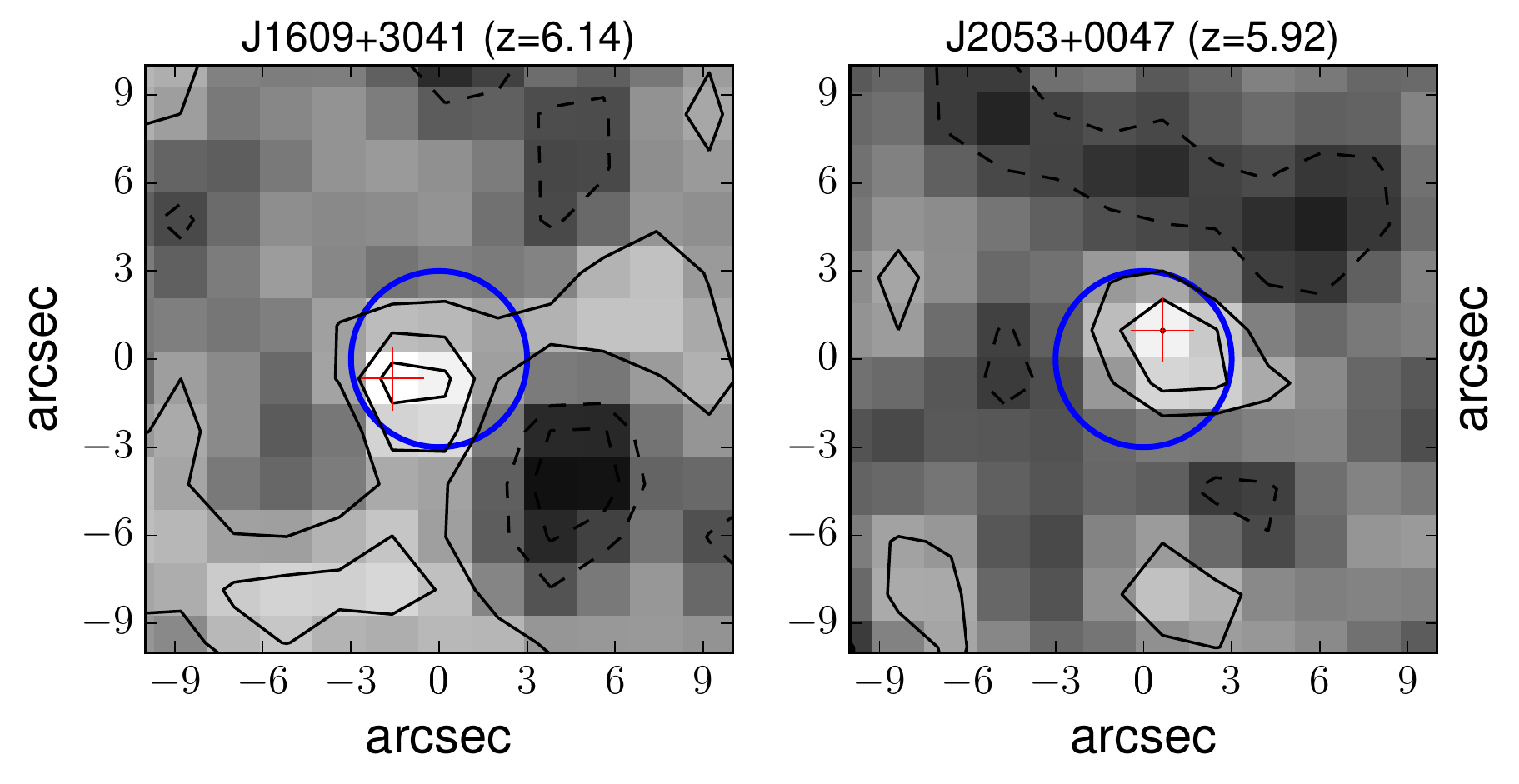}
 \caption{1.4\,GHz postage stamp images from the FIRST Survey (typical resolution of 5\arcsec) of the
 quasars J1609+3041 (left) and J2053+0047 (right). 
 The blue circles have a radius of 3\arcsec\ and
 are centered on the quasar optical positions. These positions have J2000 coordinates corresponding to (R.A., Decl.) = (16:09:37.28, +30:41:47.7) for J1609+3041,
 and (R.A., Decl.) = (20:53:21.77, +00:47:06.8) for J2053+0047.
 The red crosses show the 1.4\,GHz peak pixel. The black solid (dashed)
lines are the positive (negative) $1,2,3\sigma$ -contours. J1609+3041 and J2053+0047 have a S/N of 3.5 and 3.0, respectively. 
Using these radio flux densities, both quasars are classified as radio-loud with $R=28.3\pm 8.6$ and $R=44.1\pm18.7$, respectively (see Table~\ref{tab:literature}).} 
 \label{fig:firstqsos}
 \end{center}
 \end{figure*}

Another common definition for radio-loudness in the literature (besides our adopted criteria in Section~\ref{sec:RLD}), is a simple
cut on rest-frame radio luminosity. This criteria is a better indicator of radio-loudness if the optical and radio luminosities
are not correlated \citep[][Appendix C]{peacock1986, miller1990, ivezic2002}.
We here explore how our results would change if we adopt a fixed radio luminosity as a boundary between radio-quiet and radio-loud objects.
We use the alternative criteria adopted by \cite{jiang2007}, where a radio-loud quasar is defined with a luminosity density at rest-frame 5\,GHz,
$L_{\nu,\rm 5Ghz} > 10^{32.5}$\,ergs\,s$^{-1}$\,Hz$^{-1}$. This is equivalent to 
requiring $\log L_{\rm 5\,Ghz}(L_\sun) > 8.61 $ (see the horizonal line in Figure~\ref{fig:radio}). 

{One caveat with this definition is that J2228+0110 is just below the radio-loud cut although is still consistent
with being radio-loud within the uncertainties: $\log L_{\rm 5Ghz}(L_\sun) = 8.59\pm0.08$. Note that the quasar J0203+0012, 
discussed in Section~\ref{sec:j0203}, is also classified as radio-quiet by this definition.
}

{We estimate the radio-loud fraction using the Kaplan--Meier estimator, following the approach of the previous section.
The estimated RLF with this definition is $6.6^{+4.1}_{-1.6}\%$. This result agrees with the one obtained in
Section ~\ref{sec:rlfz6}, and they are both consistent with no strong evolution of the 
radio-loud fraction of quasars with redshift.
}

\section{Summary}
\label{sec:summary}
We perform a search for high-redshift, radio-loud quasars ($i$- and $z$-dropouts) by combining radio and optical observations 
from the FIRST and \PS\ surveys. 
The multiwavelength information of these surveys allows the identification of quasars with optical colors similar to the more 
numerous cool dwarfs and therefore missed by typical color selection used by high-redshift quasar surveys \citep[e.g.,][]{fan2006b, banados2014}.
We do not find good quasar candidates at  $z\gtrsim 6.4$ ($z$-dropouts). 
We discover two of the radio loudest quasars at $z\gtrsim 5.6$:  PSO~J055.4244--00.8035 ($z=5.68$) with a radio-loudness parameter $R=178.0 \pm 40.5$ and
PSO~J135.3860+16.2518 ($z=5.63$) with $R=91.4\pm 8.8$. 
{These two quasars are at the low-redshift end of the $i$-dropout selection technique ($5.5 \lesssim z \lesssim 5.7$) and they are
too blue in $i-z$ to have been selected by color cuts usually applied in optical searches for high-redshift quasars. Currently, there is an apparent lack of quasars at $5.2<z<5.7$
\citep[see][their Figure 4]{mcgreer2013, banados2014}. This is due to the 
similarity between optical colors of quasars and M dwarfs which are the most numerous stars in the Galaxy \citep{rojas-ayala2014}.
The identification of these two radio-loud quasars in our extended selection criteria implies that there must be a significant number of radio-quiet 
quasars at these redshifts that are just being missed by standard selection criteria. The use of additional wavelength information, for example
using \textit{WISE} photometry, might help to find quasars in this still unexplored redshift range.
}

We inspect all the 1.4\,GHz FIRST images of the known quasars that are in the FIRST survey footprint but not in the catalog.
Based on this inspection, we identify two additional $z\sim 6$ radio-loud quasars which are detected at S/N\,$\gtrsim 3$ and therefore
 would benefit from deeper radio imaging.

We highlight the importance of infrared photometry (e.g., from \textit{Spitzer} or \textit{WISE}) for $z>5.5$ quasars in order to have
an accurate measurement of the rest-frame 4400\,\AA\ luminosity which allows us to robustly classify quasars as radio-loud or radio-quiet.
By using \textit{Spitzer} photometry we reclassify the quasar J0203+0012 at $z=5.72$ as radio-quiet ($R=4.3\pm0.5$). This quasar was previously classified as
radio-loud by estimating its rest frame 4400\,\AA\ luminosity from the magnitude at 1450\,\AA\ \citep{wang2008b}.
The estimate based on an infrared proxy is much better than the one based on the $m_{1450}$ proxy because less extrapolation is needed. 
Thus, the estimated rest-frame 4400\,\AA\ luminosity is less affected
by spectral energy distribution assumptions.

We compile all the $z>5.5$ quasars having 1.4\,GHz data in the literature and, by making simple assumptions (see Section \ref{sec:rlfz6}), 
we find that the radio-loud fraction of quasars at $z\sim 6$ is between 6\% and 19\%.
{We also estimate the radio-loud fraction using the Kaplan--Meier estimator which takes into account both radio detections and upper limits,
obtaining RLF$\, =8.1^{+5.0}_{-3.2}\%$.
This fraction suggests no strong evolution of the radio-loud fraction with redshift. This result contrasts
with some lower redshift studies that show a decrease of the radio-loud fraction of quasars with
redshift  \citep[e.g.,][]{jiang2007}.
}

The study of the RLF of quasars at $z \gtrsim 6$ has the potential to give a definitive answer to the issue of a possible evolution of
the RLF of quasars with redshift. For this, a homogeneous radio (and infrared) follow-up of a well-defined sample of $z\sim 6$ quasars
(or $z\sim 7$ when more of these objects are discovered) selected in a consistent manner is crucial to test whether there is evolution 
in the RLF of quasars.

\acknowledgments
We thank the anonymous referee for providing excellent suggestions and comments which improved the manuscript.

EB thanks the IMPRS for Astronomy \& Cosmic Physics at the University of Heidelberg.

EPF and BPV acknowledge funding through the ERC grant `Cosmic Dawn'.

We thank F.~Ardila, M. Balokovi\'c, J. Larson, E. Manjavacas, M.~Maseda,  T.~Minear, A. Place, S.~Schmidl, and C.~Steinhardt for their important
participation in some of our follow-up observations.

The Pan-STARRS1 Surveys (PS1) have been made possible through contributions of the Institute for Astronomy, the University of Hawaii, the Pan-STARRS Project Office, the Max-Planck Society and its participating institutes, the Max Planck Institute for Astronomy, Heidelberg and the Max Planck Institute for Extraterrestrial Physics, Garching, The Johns Hopkins University, Durham University, the University of Edinburgh, Queen's University Belfast, the Harvard-Smithsonian Center for Astrophysics, the Las Cumbres Observatory Global Telescope Network Incorporated, the National Central University of Taiwan, the Space Telescope Science Institute, the National Aeronautics and Space Administration under grant No. NNX08AR22G issued through the Planetary Science Division of the NASA Science Mission Directorate, the National Science Foundation under grant No. AST-1238877, the University of Maryland, and Eotvos Lorand University (ELTE).

This work is based on observations made with ESO Telescopes at the La Silla Paranal Observatory under programs
ID 092.A-0150, 093.A-0863, and 093.A-0574.

The LBT is an international collaboration among institutions in the United States, Italy and Germany. The LBT Corporation partners are: The University of Arizona on behalf of the Arizona university system; Istituto Nazionale di Astrofisica, Italy;  LBT Beteiligungsgesellschaft, Germany, representing the Max Planck Society, the Astrophysical Institute Potsdam, and Heidelberg University; The Ohio State University; The Research Corporation, on behalf of The University of Notre Dame, University of Minnesota and University of Virginia.

 This paper used data obtained with the MODS spectrographs built with
 funding from NSF grant AST-9987045 and the NSF Telescope System
 Instrumentation Program (TSIP), with additional funds from the Ohio
 Board of Regents and the Ohio State University Office of Research.

Part of the funding for GROND (both hardware as well as personnel) was 
  generously granted from the Leibniz-Prize to Prof. G. Hasinger 
  (DFG grant HA 1850/28-1).

This publication makes use of data products from the \textit{Wide-field Infrared Survey Explorer},
which is a joint project of the University of California, Los Angeles, and the Jet Propulsion Laboratory/California Institute of Technology,
funded by the National Aeronautics and Space Administration.

We used the Milliquas Quasar Catalog to cross match our candidates with known quasars (\url{http://quasars.org/milliquas.htm}; \citealt{flesch2015})
This research made use of Astropy, a community-developed core Python package for Astronomy \citep[][\url{http://www.astropy.org}]{astropy13}.
We used the python package Lifelines \citep[][\url{https://github.com/camdavidsonpilon/lifelines}]{lifelines} to perform the Kaplan-Meier estimates.
This publication made use of TOPCAT  \citep[][\url{http://www.starlink.ac.uk/topcat}]{taylor2005}.
The plots in this publication 
were produced using Matplotlib \citep[][\url{http://www.matplotlib.org}]{hunter2007}.

{\it Facilities:} \facility{PS1 (GPC1)},\facility{VLT:Antu (FORS2)}, \facility{NTT (EFOSC2)}, \facility{LBT (MODS)}, \facility{Max Planck:2.2m (GROND)}
, \facility{Hale (DBSP)}



\begin{turnpage}
\input{t1deluxe}
\end{turnpage}
\global\pdfpageattr\expandafter{\the\pdfpageattr/Rotate 90}

\begin{turnpage}

\input{t2deluxe}
\end{turnpage}

\clearpage

\input{t3deluxe}
\global\pdfpageattr\expandafter{\the\pdfpageattr/Rotate 0}



\end{document}

%% file: t1deluxe.tex
\begin{deluxetable*}{lcccccccccc}
\tablecolumns{12}
\tabletypesize{\scriptsize}
\tablewidth{0pc}
\tablecaption{Candidates After Visual Inspection \label{tab:candidates}}
 \tablehead{
\colhead{Candidate}    &  \colhead{ R.A. } &   \colhead{Decl.}   & \colhead{\gps } &
\colhead{\rps } & \colhead{\ips }   & \colhead{\zps }  
& \colhead{\yps } & \colhead{\speak }    & \colhead{Prio. \tablenotemark{a}  }  
\\
\colhead{}    &  \colhead{(J2000)} &   \colhead{(J2000)}   & \colhead{(mag)} &
\colhead{(mag)} & \colhead{(mag)}   & \colhead{(mag)}  
& \colhead{(mag)} & \colhead{(mJy)}    & \colhead{} 
}
\startdata
PSO J141.7159+59.5142  & 09:26:51.83	& +59:30:51.2  & $>23.69$       & $>23.39$        &$>23.27$           & $>22.71$        &  $20.64\pm 0.14$ & $1.11 \pm 0.15$  & L  \\
PSO J172.3556+18.7734  & 11:29:25.36	& +18:46:24.3  & $>23.64$       & $>23.41$        &$>23.26$           & $>22.78$        &  $20.68\pm 0.16$ & $1.02 \pm 0.14$   & L  \\
\hline
PSO J044.9329--02.9977 & 02:59:43.91    & --02:59:51.9 & $>22.81$       & $>23.33$        &$22.88 \pm 0.28$   & $21.16\pm0.08$  &  $20.89\pm 0.17$ & $5.86 \pm 0.08$  & H  \\
PSO J049.0958--06.8564 & 03:16:23.00    & --06:51:23.2 & $>22.57$       & $>23.11$        &$23.07 \pm 0.28$   & $21.37\pm0.10$  &  $20.76\pm 0.17$ & $2.77 \pm 0.14$  & H  \\
PSO J055.4244--00.8035 & 03:41:41.86    & --00:48:12.7 & $>22.88$       & $>23.08$        &$>21.54$           & $20.28\pm0.05$  &  $20.27\pm 0.10$ & $2.14 \pm 0.14$  & H  \\
PSO J106.7475+40.4145  & 07:06:59.40	& +40:24:52.3  & $>23.57$       & $>23.44$        &$>22.73$           & $21.39\pm0.10$  &  $20.85\pm 0.17$ & $1.37\pm0.13$  & L  \\
PSO J114.6345+25.6724  & 07:38:32.30	& +25:40:20.8  & $>23.55$       & $22.95\pm 0.14$ &$22.48 \pm 0.25$   & $20.89\pm0.10$  &  $20.62\pm 0.15$ & $ 6.75\pm0.13 $  & H  \\
PSO J135.3860+16.2518  & 09:01:32.65    & +16:15:06.8  & $23.61\pm0.24$ & $>23.97 $       &$22.38 \pm 0.17$   & $20.67\pm0.05$  &  $20.82\pm 0.15$ & $3.04 \pm 0.14$  & H  \\
PSO J164.9800+07.4459  & 10:59:55.22	& +07:26:45.5  & $>23.08$       & $>22.68$        &$21.71 \pm 0.14$   & $20.17\pm0.05$  &  $>21.39$ & $ 3.35\pm0.14 $  & H  \\
PSO J208.4897+11.8071  & 13:53:57.54	& +11:48:25.6  & $>23.46$       & $23.27\pm 0.28$ &$22.73 \pm 0.15$   & $20.96\pm0.09$  &  $20.79 \pm 0.14$ & $ 2.28\pm0.13 $  & H  \\
PSO J238.0370--03.5494 & 15:52:08.89    & --03:32:58.0 & $>23.57$       & $>23.6 $        &$22.85 \pm 0.33$   & $21.15\pm 0.07$ &  $20.84 \pm 0.19$ & $ 6.01\pm0.15 $  & H  \\
PSO J354.6110+04.9453  & 23:38:26.65	& +04:56:43.3  & $>23.46$       &$23.01\pm0.19 $ &$22.77 \pm0.18 $  & $20.88\pm0.09 $ &  $20.98 \pm 0.29 $ & $ 6.44\pm0.13 $  & H  \\
\enddata
\tablecomments{The two entries at the top are $z$-dropouts and the ten at the bottom are $i$-dropouts. The lower limits correspond to $3\sigma$ limiting magnitudes.}
\tablenotetext{a}{Priorities. H: High. L: Low}

\end{deluxetable*}



%% file: t2deluxe.tex
\begin{deluxetable*}{lccccccccccc}
\tablecolumns{12}
\tabletypesize{\scriptsize}
\tablewidth{0pc}
\tablecaption{High Priority $i$-dropout Candidates Follow-up.  \label{tab:followup} }
 \tablehead{
\colhead{Candidate}    &  \colhead{ \ggrond} &   \colhead{\rgrond}   & \colhead{\igrond} &
\colhead{\zgrond} & \colhead{\Jgrond}   & \colhead{\Hgrond}  
& \colhead{\Kgrond} & \colhead{\intt }    & \colhead{\zntt }  & \colhead{\Jntt }  & \colhead{Note \tablenotemark{a} } 
\\
\colhead{}    &  \colhead{(mag)} &   \colhead{(mag)}   & \colhead{(mag)} &
\colhead{(mag)} & \colhead{(mag)}   & \colhead{(mag)}  
& \colhead{(mag)} & \colhead{(mag)}    & \colhead{(mag)}  & \colhead{(mag)}  & \colhead{} 
}
\startdata
PSO J044.9329--02.9977 &  $>23.72$       & $23.37\pm0.29$ & $22.20\pm0.18$ & $22.14\pm0.23$ & $20.35\pm0.17$ & $19.92\pm0.17$ & $19.20 \pm0.28$ & --             & --             & --             & 3 \\
PSO J049.0958--06.8564 &  $24.47\pm0.33$ & $23.90\pm0.23$ & $23.12\pm0.26$ & $21.98\pm0.12$ & $20.52\pm0.15$ & $20.01\pm0.18$ & $19.19 \pm0.24$ & --             & --             & --             & 3 \\
PSO J055.4244--00.8035 &  $>23.73$       & $>23.77$       & $22.16\pm0.18$ & $20.58\pm0.05$ & $20.08\pm0.16$ & $20.03\pm0.22$ & $>19.09$        & --             & --             & --             & 1 \\
PSO J114.6345+25.6724  &  --             &  --            & --             & --             & --             & --             & --              & $22.04\pm0.18$ & $20.95\pm0.28$ & $20.48\pm0.11$ & 2 \\
PSO J135.3860+16.2518  &  $>24.57$       & $24.32\pm0.36$ & $22.70\pm0.18$ & $20.85\pm0.04$ & $20.30\pm0.12$ & $20.91\pm0.33$ & $>19.71$        & --             & --             & --             & 1 \\
PSO J164.9800+07.4459  &  $23.66\pm0.15$ & $22.19\pm0.06$ & $21.15\pm0.05$ & $20.49\pm0.02$ & $19.96\pm0.10$ & $19.72\pm0.13$ & $19.45 \pm0.27$ & $20.64\pm0.04$ & $20.27\pm 0.04$& --             & 3 \\
PSO J208.4897+11.8071  &  $>24.50$       & $23.21\pm0.18$ & $22.08\pm0.25$ & $21.66\pm0.11$ & $20.53\pm0.23$ & $19.83\pm0.19$ & $>19.46$        & --             & --             & --             & 3 \\
PSO J238.0370--03.5494 &  $>24.47$       & $23.44\pm0.18$ & $22.03\pm0.11$ & $21.32\pm0.07$ & $19.99\pm0.11$ & $19.74\pm0.16$ & $>19.70$        & --             & --             & --             & 3 \\
PSO J354.6110+04.9453  &  --             &  --            & --             & --             & --             & --             & --              & --             & --             & --             & 2 \\
\enddata
\tablecomments{The magnitude lower limits correspond to $3\sigma$ limiting magnitudes.}
\tablenotetext{a}{1: $z>5.5$ quasar spectroscopically confirmed in this work. 2: Not a $z>5.5$ quasar based on follow-up spectroscopy. 
3: Not a $z>5.5$ quasar based on follow-up photometry.}

\end{deluxetable*}



%% file: t3deluxe.tex
\begin{deluxetable*}{cccccccccc}
\tablecolumns{10}
\tabletypesize{\footnotesize}
\tablewidth{0pc}
\tablecaption{Data and Derived Parameters of the $z>5.5$ Quasars with 1.4GHz Data in the Literature. \label{tab:literature}}
 \tablehead{
\colhead{Quasar}    &  \colhead{ $z$} &   \colhead{\speak}   & \colhead{Ref.($z$,1.4\,GHz)} &
\colhead{$m_{1450}$ \tablenotemark{a}} & \colhead{$W_1$ \tablenotemark{b}}   & \colhead{$S_{3.6\mu \rm{m}}$\tablenotemark{c} }  
& \colhead{$\log L_{5\,{\rm GHz}}$} & \colhead{$\log L_{4400\angstrom}\tablenotemark{d}$}    & \colhead{$R$ \tablenotemark{d}} 
\\
\colhead{}    &  \colhead{} &   \colhead{($\mu \rm{Jy}$)}   & \colhead{} &
\colhead{(mag)} & \colhead{(mag)}   & \colhead{(mag)}  
& \colhead{($L_{\sun}$)} & \colhead{($L_{\sun}$)}    & \colhead{} 
}
\startdata
J0002+2550 & 5.82 & $89 \pm 14$ & 1,2 & 19.0 & $18.86 \pm 0.06$ & $18.71 \pm 0.02$ & $8.03 \pm 0.07$ & $13.05 \pm 0.01$ & $1.3 \pm 0.2$ \\
J0005--0006 & 5.85 & $<390$ & 3,4 & 20.2 & $20.00 \pm 0.16$ & $20.10 \pm 0.03$ & $<8.7$ & $12.49 \pm 0.01$ & $<20.9$ \\
J0033--0125 & 6.13 & $<57$ & 5,4 & 21.8 & $20.99 \pm 0.40$ & -- & $<7.9$ & $12.19 \pm 0.15$ & $<6.8$ \\
J0203+0012 & 5.72 & $195 \pm 22$ & 6,4 & 21.0 & $19.39 \pm 0.09$ & $19.14 \pm 0.03$ & $8.36 \pm 0.05$ & $12.86 \pm 0.01$ & $4.3 \pm 0.5$ \\
J0303--0019 & 6.08 & $<186$ & 7,4 & 21.3 & -- & $20.24 \pm 0.04$ & $<8.4$ & $12.47 \pm 0.01$ & $<11.4$ \\
PJ055--00 & 5.68 & $2140 \pm 137$ & 8,9 & 20.4 & $20.62 \pm 0.26$ & -- & $9.39 \pm 0.03$ & $12.27 \pm 0.09$ & $178.0 \pm 40.5$ \\
J0353+0104 & 6.049 & $<57$ & 10,4 & 20.2 & $19.34 \pm 0.09$ & $19.44 \pm 0.04$ & $<7.9$ & $12.79 \pm 0.01$ & $<1.7$ \\
J0818+1722 & 6.02 & $123 \pm 12$ & 1,2 & 19.3 & -- & $18.35 \pm 0.01$ & $8.20 \pm 0.04$ & $13.22 \pm 0.01$ & $1.3 \pm 0.1$ \\
J0836+0054 & 5.81 & $1740 \pm 40$ & 3,2 & 18.8 & $18.00 \pm 0.04$ & $17.87 \pm 0.01$ & $9.32 \pm 0.01$ & $13.38 \pm 0.01$ & $11.9 \pm 0.3$ \\
J0840+5624 & 5.8441 & $<27$ & 11,2 & 20.0 & $19.46 \pm 0.14$ & $19.53 \pm 0.02$ & $<7.5$ & $12.72 \pm 0.01$ & $<0.9$ \\
J0841+2905 & 5.98 & $<81$ & 1,4 & 19.6 & $19.91 \pm 0.16$ & $19.74 \pm 0.05$ & $<8.0$ & $12.65 \pm 0.02$ & $<3.1$ \\
J0842+1218 & 6.08 & $<57$ & 12,4 & 19.6 & -- & $19.13 \pm 0.01$ & $<7.9$ & $12.91 \pm 0.01$ & $<1.3$ \\
PJ135+16 & 5.63 & $3040 \pm 145$ & 8,9 & 20.6 & $19.51 \pm 0.11$ & -- & $9.53 \pm 0.02$ & $12.71 \pm 0.04$ & $91.4 \pm 8.8$ \\
J0927+2001 & 5.7722 & $50 \pm 11$ & 13,4 & 19.9 & $19.40 \pm 0.11$ & $19.72 \pm 0.05$ & $7.77 \pm 0.10$ & $12.63 \pm 0.02$ & $1.9 \pm 0.4$ \\
J1030+0524 & 6.308 & $<60$ & 3,2 & 19.7 & $19.28 \pm 0.09$ & $19.23 \pm 0.04$ & $<7.9$ & $12.91 \pm 0.02$ & $<1.5$ \\
J1044--0125 & 5.7847 & $<72$ & 14,2 & 19.2 & $19.05 \pm 0.07$ & $18.84 \pm 0.02$ & $<7.9$ & $12.99 \pm 0.01$ & $<1.2$ \\
J1048+4637 & 6.2284 & $<33$ & 1,2 & 19.2 & $19.05 \pm 0.06$ & $18.80 \pm 0.01$ & $<7.7$ & $13.07 \pm 0.01$ & $<0.5$ \\
J1120+0641 & 7.0842 & $<23$ & 15,16 & 20.4 & $19.61 \pm 0.11$ & $19.39 \pm 0.03$ & $<7.6$ & $12.93 \pm 0.01$ & $<0.7$ \\
J1137+3549 & 6.03 & $<51$ & 1,2 & 19.6 & $19.16 \pm 0.07$ & $19.09 \pm 0.03$ & $<7.8$ & $12.92 \pm 0.01$ & $<1.1$ \\
J1148+5251 & 6.4189 & $55 \pm 12$ & 1,17 & 19.0 & $18.67 \pm 0.05$ & $18.57 \pm 0.02$ & $7.91 \pm 0.09$ & $13.18 \pm 0.01$ & $0.7 \pm 0.2$ \\
J1250+3130 & 6.15 & $<63$ & 1,2 & 19.6 & $19.11 \pm 0.07$ & $19.09 \pm 0.01$ & $<7.9$ & $12.94 \pm 0.01$ & $<1.3$ \\
J1306+0356 & 6.016 & $<63$ & 3,2 & 19.6 & $19.57 \pm 0.10$ & $19.24 \pm 0.04$ & $<7.9$ & $12.86 \pm 0.02$ & $<1.5$ \\
J1319+0950 & 6.133 & $64 \pm 17$ & 14,18 & 19.6 & $19.73 \pm 0.11$ & -- & $7.94 \pm 0.12$ & $12.69 \pm 0.04$ & $2.4 \pm 0.7$ \\
J1335+3533 & 5.9012 & $35 \pm 10$ & 1,2 & 19.9 & $19.41 \pm 0.07$ & $19.35 \pm 0.02$ & $7.64 \pm 0.12$ & $12.80 \pm 0.01$ & $0.9 \pm 0.3$ \\
J1411+1217 & 5.904 & $61 \pm 16$ & 3,2 & 20.0 & $19.29 \pm 0.07$ & $19.05 \pm 0.02$ & $7.88 \pm 0.11$ & $12.92 \pm 0.01$ & $1.2 \pm 0.3$ \\
J1425+3254\tablenotemark{e} & 5.8918 & $<60$ & 1,4 & 20.6 & $19.67 \pm 0.08$\tablenotemark{e} & $20.36 \pm 0.06$\tablenotemark{e}& $<7.9$ & $12.39 \pm 0.02$ & $<4.1$\tablenotemark{e} \\
J1427+3312 & 6.12 & $1730 \pm 131$ & 19,9 & 20.3 & $19.52 \pm 0.08$ & $19.49 \pm 0.02$ & $9.37 \pm 0.03$ & $12.77 \pm 0.01$ & $53.3 \pm 4.1$ \\
J1429+5447 & 6.1831 & $2930 \pm 152$ & 18,9 & 20.9 & $19.73 \pm 0.08$ & -- & $9.60 \pm 0.02$ & $12.70 \pm 0.03$ & $109.2 \pm 8.9$ \\
J1436+5007 & 5.85 & $<48$ & 1,2 & 20.2 & $19.87 \pm 0.09$ & $19.79 \pm 0.02$ & $<7.8$ & $12.62 \pm 0.01$ & $<1.9$ \\
J1509--1749 & 6.121 & $<54$ & 20,18 & 19.8 & -- & -- & $<7.9$ & $12.91 \pm 0.11$ & $<1.2$ \\
J1602+4228 & 6.09 & $60 \pm 15$ & 1,2 & 19.9 & $18.75 \pm 0.04$ & $18.57 \pm 0.02$ & $7.90 \pm 0.11$ & $13.14 \pm 0.01$ & $0.8 \pm 0.2$ \\
J1609+3041 & 6.14 & $484 \pm 137$ & 21,22 & 20.9 & $20.22 \pm 0.14$ & -- & $8.82 \pm 0.12$ & $12.50 \pm 0.05$ & $28.3 \pm 8.6$ \\
J1621+5155 & 5.71 & $<63$ & 4,4 & 19.9 & $18.35 \pm 0.03$ & -- & $<7.9$ & $13.18 \pm 0.01$ & $<0.7$ \\
J1623+3112 & 6.26 & $<93$ & 18,2 & 20.1 & $19.22 \pm 0.06$ & $19.23 \pm 0.03$ & $<8.1$ & $12.90 \pm 0.01$ & $<2.3$ \\
J1630+4012 & 6.065 & $<45$ & 1,4 & 20.6 & $20.19 \pm 0.12$ & $19.98 \pm 0.06$ & $<7.8$ & $12.57 \pm 0.02$ & $<2.2$ \\
J1641+3755 & 6.047 & $<96$ & 20,4 & 20.6 & -- & -- & $<8.1$ & $12.58 \pm 0.12$ & $<4.6$ \\
J2053+0047 & 5.92 & $434 \pm 143$ & 23,22 & 21.2 & $20.82 \pm 0.32$ & -- & $8.74 \pm 0.14$ & $12.23 \pm 0.12$ & $44.1 \pm 18.7$ \\
J2054--0005 & 6.0391 & $<69$ & 14,4 & 20.6 & -- & -- & $<8.0$ & $12.59 \pm 0.12$ & $<3.2$ \\
J2147+0107 & 5.81 & $<54$ & 23,18 & 21.6 & $20.33 \pm 0.20$ & -- & $<7.8$ & $12.41 \pm 0.07$ & $<3.5$ \\
J2228+0110 & 5.95 & $310 \pm 57$ & 24,24 & 22.2 & -- & -- & $8.59 \pm 0.08$ & $11.94 \pm 0.13$ & $61.3 \pm 20.9$ \\
J2307+0031 & 5.87 & $<51$ & 23,18 & 21.7 & $19.78 \pm 0.13$ & -- & $<7.8$ & $12.64 \pm 0.05$ & $<2.0$ \\
J2315--0023 & 6.117 & $<48$ & 10,4 & 21.3 & $20.26 \pm 0.20$ & $20.10 \pm 0.03$ & $<7.8$ & $12.53 \pm 0.01$ & $<2.6$ \\
J2329--0301 & 6.417 & $<66$ & 20,4 & 21.6 & -- & -- & $<8.0$ & $12.21 \pm 0.12$ & $<8.3$ \\
\enddata 
\tablerefs{(1) \cite{carilli2010}, (2) \cite{wang2007}, (3) \cite{kurk2007}, (4) \cite{wang2008b}, (5) \cite{willott2007}, (6) \cite{mortlock2009}, (7) \cite{kurk2009}, (8) This Work, (9) FIRST \cite{becker1995}, (10) \cite{jiang2008}, (11) \cite{wang2010}, (12) \cite{derosa2011}, (13) \cite{carilli2007}, (14) \cite{wang2013}, (15) \cite{venemans2012}, (16) \cite{momjian2014}, (17) \cite{carilli2004}, (18) \cite{wang2011}, (19) \cite{mcgreer2006}, (20) \cite{willott2010b}, (21) Warren et al. (in prep.), (22) This Work: new radio identification (see Section \ref{sec:newfirst}), (23) \cite{jiang2009}, (24) \cite{zeimann2011}}
\tablecomments{ Reported upper limits correspond to $3\sigma$.}
\tablenotetext{a}{All $m_{1450}$ are taken from \cite{calura2014}
              except for J2228+0110 for which is taken from \cite{zeimann2011},
              for J1609+3041 for which is calculated from its \yps\ band (Ba\~{n}ados et al. in prep.), and for 
              PJ055-00 and PJ135+16 for which are calculated in this work
              as in \cite{banados2014}.}
\tablenotetext{b}{All $W_1$ measurements have $S/N>2.5$.
                Magnitudes are taken from the main ALLWISE source catalog with 
                exception of J0033-0125, PJ055-00, and J2053+0047, for which are taken
                from the ALLWISE reject table.}
\tablenotetext{c}{$S_{3.6\mu \rm{m}}$ measurements are from \cite{leipski2014} with exception of J1120+0641 and J1425+3254 for which are taken from \cite{barnett2015} and \cite{cool2006}, respectively.}
\tablenotetext{d}{$\log L_{4400\angstrom}$ and $f_{\nu,4400 \angstrom}$ in $R=f_{\nu,5 \,{\rm GHz}} / f_{\nu,4400 \angstrom}$
 are based on $S_{3.6\mu \rm{m}}$ measurements when available, otherwise from $W_1$.
 If the quasar does not have $S_{3.6\mu \rm{m}}$ nor $W_1$ data, the quantities are extrapolated
 from $m_{1450}$ (see text in Section \ref{sec:optical}).}
 \tablenotetext{e}{We note a large discrepancy between the reported \textit{WISE} and \textit{Spitzer} magnitudes for J1425+3254. If $W_1$ is used instead of  $S_{3.6\mu \rm{m}}$ to estimate $f_{\nu,4400 \angstrom}$, $R$ would be $< 2.1$. }
\end{deluxetable*}